\documentclass[
 reprint,
 amsmath,amssymb,
 aps,prl,eps,superscriptaddress,twocolumn,10pt
]{revtex4}

\usepackage{graphicx}
\usepackage{dcolumn}
\usepackage{bm}
\usepackage{xcolor}

\begin{document}

\preprint{APS/123-QED}

\title{Emergence of a ferromagnetic insulating state in LaMnO$_3$/SrTiO$_3$ heterostructures: The role of strong electronic correlations and strain}

\author{Hrishit Banerjee}
\author{Markus Aichhorn}%
 \email{aichhorn@tugraz.at}
\affiliation{Institute of Theoretical and Computational Physics, Graz University of Technology, NAWI Graz, Petersga{\ss}e 16, Graz, 8010, Austria.
}

\date{\today}

\begin{abstract}
 Inspired by the experimental findings of an exotic ferromagnetic insulating state in LaMnO$_3$/SrTiO$_3$ heterostructures, we calculate the electronic and magnetic state of LaMnO$_3$/SrTiO$_3$ superlattices with comparable thicknesses employing \textit{ab-initio} dynamical mean-field theory.
 Projecting on the low-energy subspace of Mn $3d$ and Ti $3d$ states, and solving a multi-impurity problem, our approach emphasizes on local correlations at Mn and Ti sites. We find that a ferromagnetic insulating state emerges due to intrinsic effects of strong correlations in the system, in agreement with experimental studies. We also predict that, due to electronic correlations, the emerging 2D electron gas is located at the LMO side of the interface. This is in contrast to DFT results that locate the electron gas on the STO side. We estimate the transition temperature for the paramagnetic to ferromagnetic phase transition, which may be verified experimentally. Importantly, we also clarify that the epitaxial strain is a key ingredient for the emergence of the exotic ferromagnetic insulating state. This becomes clear from calculations on a strained LaMnO$_3$ system, also showing ferromagnetism which is not seen in the unstrained bulk material.
\end{abstract}

\maketitle


\paragraph{Introduction}

Since the first report of a highly conducting and mobile 2-dimensional electron gas (2DEG) occurring at the interface of oxide insulators~\cite{lao-sto}, the study of hetero-interfaces formed between perovskite oxides has made a serious impact on the scientific community engaged in experimental and computational condensed matter research. It has paved the way to many different prospective device applications, and garnered considerable interest in the field of oxide electronics. Interfaces have been formed between band insulators, for example between SrTiO$_3$ (STO) and LaAlO$_3$ (LAO)~\cite{lao-sto,sc-fm}, and between band insulators and Mott insulators, as in the case of SrTiO$_3$ and GdTiO$_3$ (GTO)~\cite{gto-sto, stemmer}, with qualitatively different behaviour from that of LAO/STO interface. This is due to the fact that GdTiO$_3$, being a Mott insulator, has a very different band structure compared to LAO which is a band insulator, which in turn has a deep influence on the band alignment of the two oxides making up the interface~\cite{banerjee-ch6}.

Experimental studies on interfaces between cooperative Jahn-Teller (JT) distortion driven insulators like LaMnO$_3$ (LMO), in which strong correlations have a significant effect, with band insulators like
SrTiO$_3$ have been carried out in the recent past. In these systems, the interplay of structural distortions and strong electronic correlations are expected to lead to a variety of different phases. Indeed, a large and diverse number of magnetic and electronic phases of the LMO/STO interfaces have been observed,
depending on the relative thickness of STO and LMO and their geometry~ \cite{barriocanal-ch6,garcia-ch6,choi-ch6,liu2-ch6,sumilan-ch6,wang2015,oor}. 
Amongst these varied number of electronic and magnetic states, the most intriguing is the observation of ferromagnetic insulating behavior, which has been reported for both LMO/STO 
superlattices and thin-film/substrate geometries when LMO and STO have comparable thicknesses~\cite{oor, sumilan-ch6}. Since ferromagnetism is generally associated with metallicity, and antiferromagnetism is typically seen in case of insulators, this happens to be a counter-intuitive observation. 

Ferromagnetic insulators are essential for many new magnetic devices, such as dissipation-less quantum-spintronic devices, magnetic tunneling junctions, etc. Ferromagnetic insulators with a high $T_\text{C}$ and a high-symmetry crystal structure are critical for integration with common single-crystalline oxide films or substrates. The few known high-symmetry materials either have extremely low Curie temperatures ($\leq 16$\,K)~\cite{mauger, katmis}, or require chemical doping of an otherwise antiferromagnetic matrix. Thus, it is imperative to theoretically understand the origin of intrinsic ferromagnetic insulating behaviour in heterostructures. 

Few attempts have been made in case of LMO/STO to theoretically justify the observed coexistence of ferromagnetism and insulating properties. Most of these attempts depend on either symmetry lowering or on orbital polarisation effects. However, symmetry lowering in the geometry of strained LMO~\cite{hou-ch6} seems unlikely in superlattices because LMO is sandwiched between cubic STO. Concerning orbital polarisation, experiments show a significant suppression of the JT distortion, further supported by DFT+U studies~\cite{spaldin-ch6, hb-kh}, which in turn reduces orbital polarisation~\cite{oor} in
the superlattice geometry that hosts the ferromagnetic insulating behavior. A further attempt suggests electronic phase separation leading to the nucleation of metallic nanoscale ferromagnetic islands embedded in an insulating antiferromagnetic matrix. This, however, is not the case of intrinsic ferromagnetic insulating behaviour~\cite{sumilan-ch6}. The observed coexistence of ferromagnetism and insulating behavior in the LMO/STO heterostructure thus remains a mystery. 

Saha-Dasgupta~\textit{et al.} studied the problem considering bulk LMO, but
epitaxially strained to the lattice constants of a square substrate of STO~\cite{hb-kh}. They consider the general framework of DFT, which is expected to capture the structural changes that happen upon epitaxial straining of LMO
correctly, supplemented with Hartree-Fock based hybrid functional calculations.
They find a ferromagnetic
ground state driven by the marked reduction of orthorhombic distortion in the optimized LMO structure,
when epitaxially strained to the square substrate of STO, resulting in a strong suppression of the JT distortion. The suppression of the JT distortion and modification of the octahedral rotation, as captured in this study, is in agreement with structural characterization of LMO/STO superlattices~\cite{oor}. The treatment of exact exchange within a hybrid functional resulted in an insulating solution. This 
result was attributed to originate from electronically driven charge disproportionation within the Mn sublattice that arises due to a strain-driven enhanced covalency between Mn and O. Calculations using hybrid functionals on a heterostructure geometry confirmed the ferromagnetic insulating result. However, charge disproportionation has not been confirmed experimentally.

The questions that we intend to address in this letter are, thus, well defined. We are interested in looking at how strong correlations affect these oxide LMO/STO heterostructure systems. In particular, we intend to examine the fate of the 2D electron gas. 
We carry out paramagnetic density-functional theory (DFT) plus dynamical mean-field theory (DMFT) calculations 
on (LMO)$_{2.5}$/(STO)$_{2.5}$ and (LMO)$_{3.5}$/(STO)$_{2.5}$ which are multi-impurity calculations including both Ti-$3d$ and Mn-$3d$, and observe a metal-to-insulator phase transition by varying the interaction strength. Furthermore, we also find a ferromagnetic insulating solution which is stable at low enough temperatures. Given the overestimation of $T_\text{C}$ in DMFT studies, we estimate the transition temperature to be in the range of $\approx 100$\,K.
Both the heterostructures, (LMO)$_{2.5}$/(STO)$_{2.5}$ and (LMO)$_{3.5}$/(STO)$_{2.5}$, yield qualitatively similar results. In contrast to previous studies, we find that the 2D electron gas generated due to the polar catastrophe moves to the LMO side of the interface and effectively dopes the Mn $3d$ orbitals. 

An important result of our study is that an exotic ferromagnetic insulating state emerges also in a calculation of a bulk LaMnO$_3$ system, which is strained by matching to a square STO substrate. As LaMnO$_3$ under ambient pressure conditions is antiferromagnetic, this opens up a way to induce phase transitions between different magnetic states as function of strain. This may be realised by modern experimental techniques of stress generation.

\paragraph{Computational Details}

Our DFT calculations for structural relaxation were carried out in a plane-wave basis  with projector-augmented wave (PAW) potentials~\cite{blochl-ch6} as implemented in the Vienna Ab-initio Simulation Package
(VASP)~\cite{kresse-ch6, kresse01-ch6}. As DFT
exchange-correlation functional we chose the generalized gradient approximation (GGA), implemented following the Perdew Burke Ernzerhof (PBE) prescription~\cite{pbe-ch6}.
For ionic relaxations, internal positions of the atoms were allowed to relax until the forces became less than 0.005\,eV/\AA. An energy cutoff of 550\,eV, and a 5$\times$5$\times$3 Monkhorst–Pack $k$-points mesh provided good convergence of the total energy.

For our DFT+DMFT calculations, we performed first DFT calculations using the PBE functional in the full-potential augmented plane-wave basis as implemented in
\textsc{wien2k}~\cite{wien2k18}. We used the largest possible muffin-tin radii and the
basis set plane-wave cutoff as defined by
${R_{\text{min}}\!\cdot\!K_{\text{max}}}$\,=\,7.5, where $R_{\text{min}}$ is the
muffin-tin radius of the oxygen atoms. The consistency between VASP and \textsc{wien2k}  results has been cross-checked. The band structure of the LMO/STO heterostructures consists of both Mn and Ti bands at the Fermi level. For paramagnetic calculations, only Mn $t_{2g}$ orbitals have been considered for the DMFT calculation, while for ferromagnetic calculations all five Mn $d$ orbitals have been taken into account to allow for high-spin solutions. In all calculations we also included the Ti $t_{2g}$ orbitals as being correlated.

We performed the DMFT calculations in a basis set of projective Wannier functions, which were calculated using the TRIQS/DFTTools package~\cite{triqs-parcollet, aichhorn1, aichhorn2, aichhorn3}. 
The Anderson impurity problems were solved using the
continuous-time quantum Monte Carlo algorithm in the hybridization expansion
(CT-HYB)~\cite{werner06} as implemented in the
TRIQS/CTHYB package~\cite{pseth-cpc}. We
performed one-shot calculations, with the double-counting correction treated in the fully-localized
limit~\cite{anisimov93}. We used the
rotationally-invariant Kanamori interaction~\cite{kanamori63}. For our calculations we used $U$ values ranging from 4.5-8\,eV and $J$ varying between 0.5-0.75\,eV to investigate the metal-to-insulator transition. As commonly used, we set the intra-orbital interaction to be $U'\!=\!U\!-\!2J$. Spin-flip and pair-hopping terms were also included in the Hamiltonian.
Real-frequency results have been obtained from imaginary-frequency data using the maximum-entropy method of analytic continuation as implemented in the TRIQS/MAXENT application~\cite{maxent}.

The problem of unstrained bulk LaMnO$_3$ has been widely studied in literature and it is well known that a large value of static effective Hubbard $U$ is required in DFT+U  methods 
to account for the experimentally observed band gap of the A-type AFM insulating state. 
For instance, a large value of effective $U=8$\,eV was required to obtain a band gap of 1.4\,eV~\cite{mellan}. A previous study shows that calculations using hybrid functionals with HF exchange of 25\% yield a value of 1.7\,eV band gap~\cite{hb-kh}. 
We want to note that the actual value of the band gap of bulk LMO is also not clear experimentally. There is a rather large distribution of values, ranging from 1.1~\cite{arima}, over 1.7~\cite{saitoh} and 1.9~\cite{jung1}, up to 2.0\,eV~\cite{jung2, kruger}. It has been observed that within DFT+U a large value of static effective $U$ is generally required to match the experimental values of the band gap.


\paragraph{LMO/STO Superlattice}

\begin{figure}
	\includegraphics[width=\columnwidth]{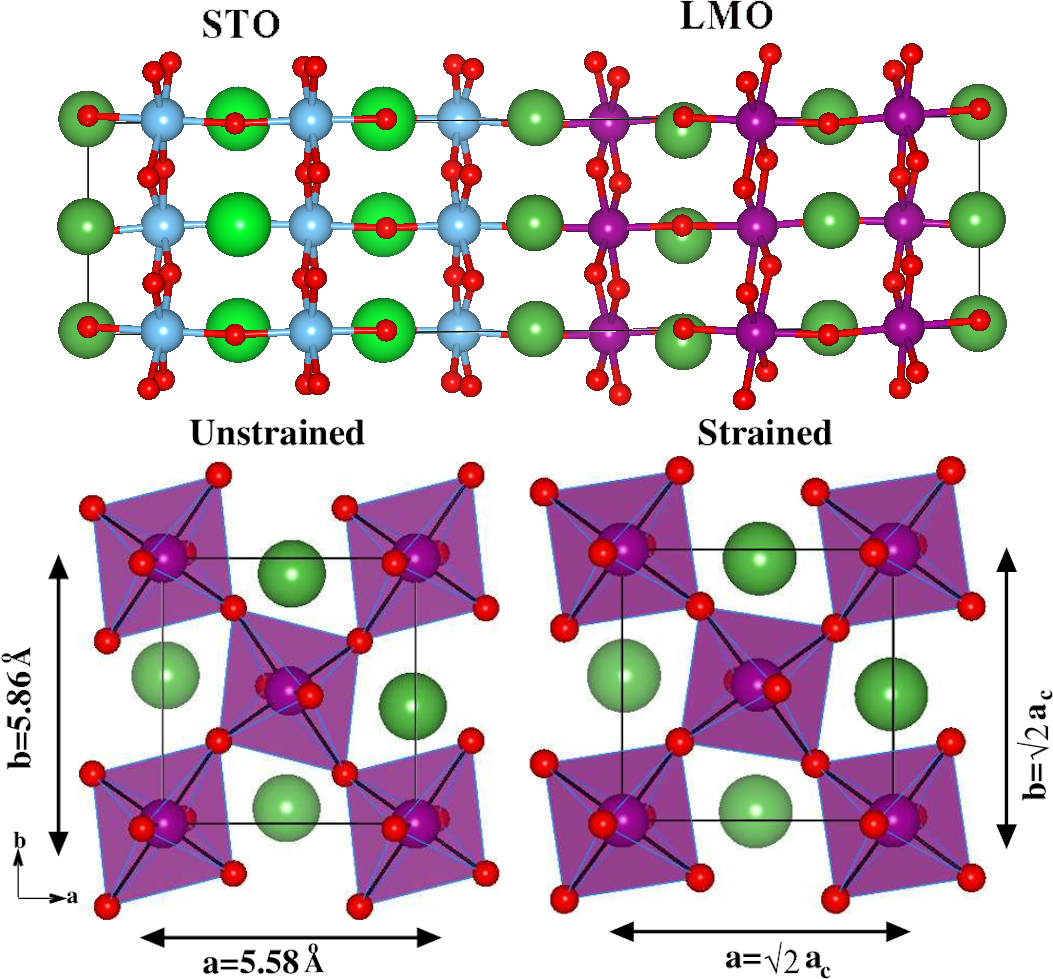}
	\caption{(Color online) Top: (LMO)$_{3.5}$/(STO)$_{2.5}$ superlattice, with 2 symmetric $n$-type interfaces. Bottomn: Unstrained and epitaxially strained LMO when the lattice constants of LMO are matched to an STO lattice.}
	\label{struct}
\end{figure}
We investigate the electronic structure of LMO on STO in the experimental set-up, i.e., we consider an actual heterostructure consisting of LMO and STO layers.
In  addition to the square epitaxial strain
generated due to the mismatch between the LMO and STO unit cells, the superlattices in particular involves the polar discontinuity formed between LMO consisting of alternating layers of LaO and MnO$_2$
of +1 and -1 charges, resp., and STO consisting of alternating charge neutral layers of SrO and TiO$_2$. The latter would cause half a charge
to be transferred between the layers at the interface. Neither the direction nor extent of this charge transfer has been clarified.
\begin{figure}
	\includegraphics[width=\columnwidth]{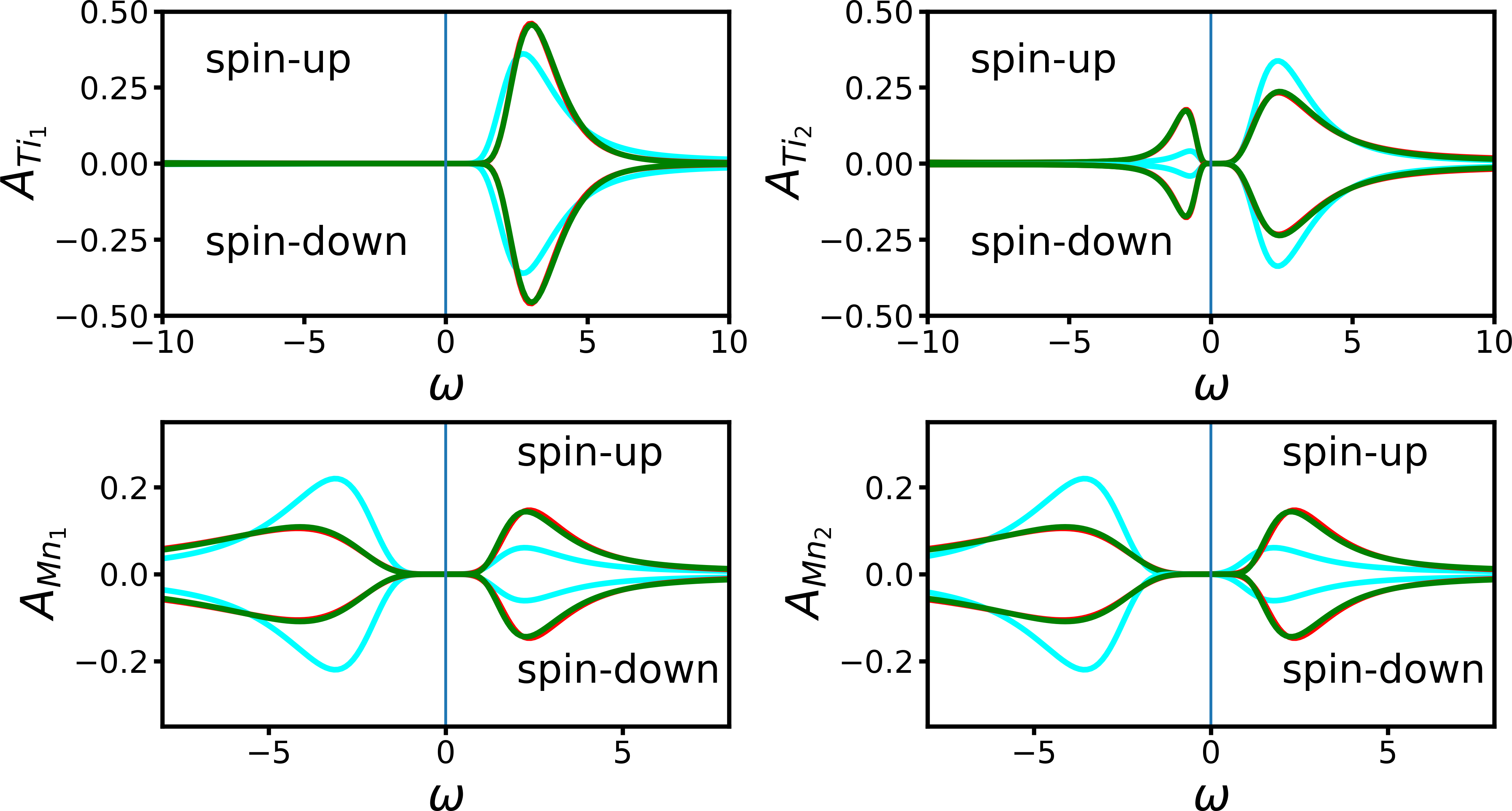}
	\caption{(Color online) Spectral functions for Ti $t_{2g}$ [$d_{xy}$(cyan), $d_{yz}$(red), $d_{xz}$ (green)] and Mn $t_{2g}$ [color code is the same as for Ti] orbitals showing the paramagnetic insulating phase at $\beta=40$\,eV$^{-1}$. Ti$_1$, Ti$_2$, Mn$_1$, Mn$_2$ belong to different inequivalent layers of the (LMO)$_{3.5}$/(STO)$_{2.5}$ superlattice and form the four impurities of the system.}
	\label{spect-par}
\end{figure}
We consider superlattice geometries of LMO/STO, with an alternate
repetition of LMO and STO layers of comparable thickness, stacked along the [001] direction. This creates two symmetric $n$-type
interfaces between the LaO layer of LMO and the TiO$_2$ layer of STO~\cite{banerjee-ch6,oor}. The structure of the superlattice for the case of (LMO)$_{3.5}$/(STO)$_{2.5}$ has been shown in the top panel of Fig. \ref{struct}. Comparable thicknesses are chosen since the
FM insulating state has been experimentally observed for superlattice geometries with nearly equal thickness of LMO and STO layers~\cite{oor}. In this letter we show multi-impurity calculations for heterostructure systems, which are not very common in existing literature.

For the structural relaxation, we place LMO in an orthorhombic geometry matching to the square plane of STO layers (in the [100] and [010] directions). Here a $\sqrt{2} \times \sqrt{2} \times c$ supercell of both LMO and STO was allowed to tilt and rotate. This
results in four Mn and Ti atoms in each MnO$_2$  and TiO$_2$ layer, respectively.
The ionic positions and $c$ lattice parameters are allowed to relax, keeping the constraint on the planar lattice constants of $a = b$. 
This generates a square-matched epitaxial strain of -1.8\%.
The optimized structure shows a significant decrease in JT distortion, and modification of tilt and rotation angles in LMO, 
while some JT distortion, and tilt and rotation is introduced in the STO due to its proximity to the distorted LMO block. This is very similar to other heterostructures such as GTO/STO~\cite{banerjee-ch6}.

\begin{figure}
	\includegraphics[width=\columnwidth]{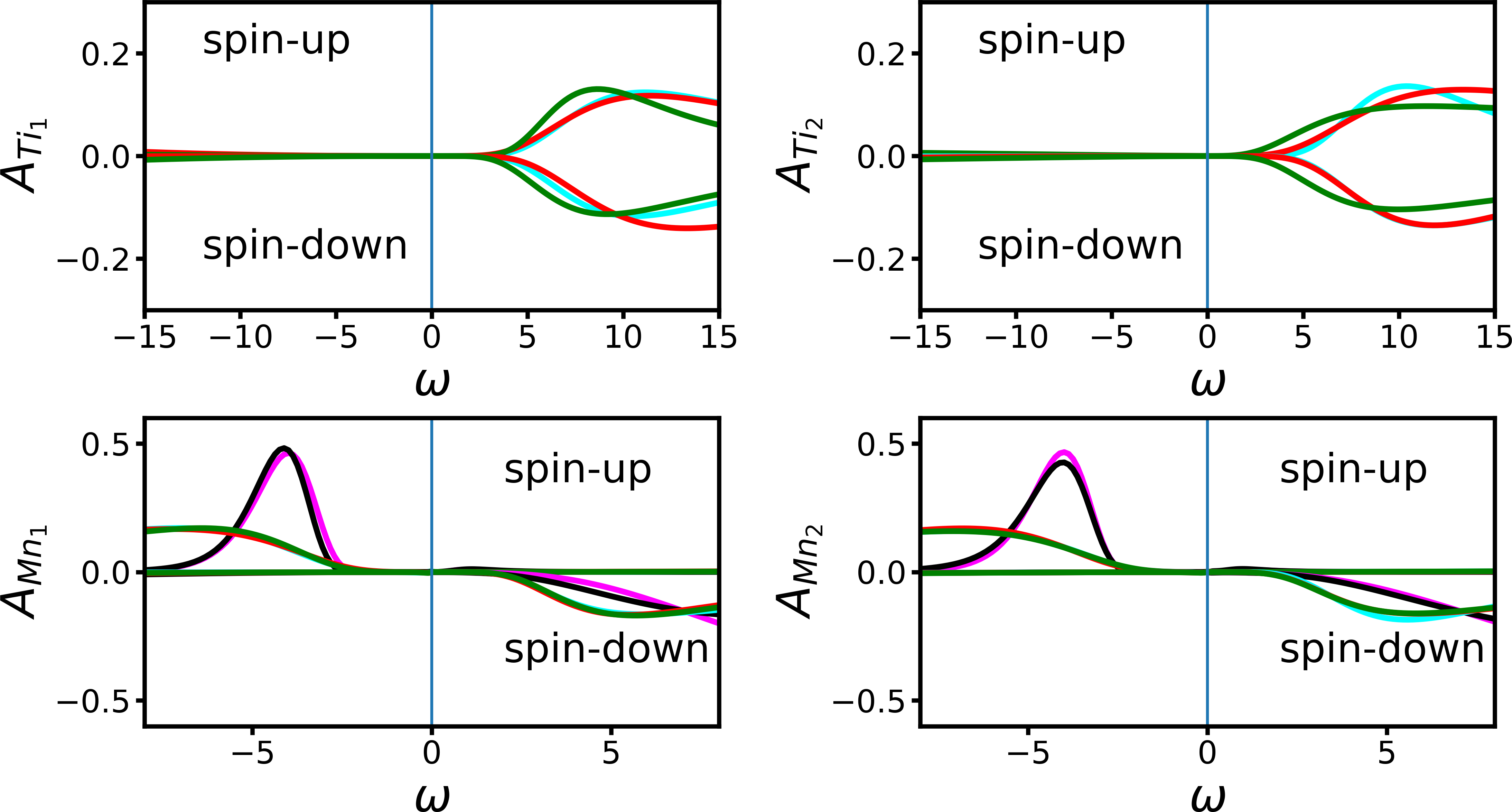}
	\caption{(Color online) Spectral functions for Ti $t_{2g}$ [$d_{xy}$(cyan), $d_{yz}$(red), $d_{xz}$ (green)] and Mn $d$[$d_{x^2-y^2}$(magenta), $d_{z^2}$ (black), and the color for $t_{2g}$ same as in case of Ti] orbitals showing FM insulating phase at $\beta=60$\,eV$^{-1}$. Ti$_1$, Ti$_2$, Mn$_1$, Mn$_2$ belong to different inequivalent layers of the (LMO)$_{3.5}$/(STO)$_{2.5}$ superlattice.}
	\label{het-fm-ins}
\end{figure}

Our paramagnetic DFT calculations reveal a metallic solution with Ti $t_{2g}$ and Mn $t_{2g}$ states at Fermi level. We first carry out paramagnetic DMFT calculations at $\beta=40$\,eV$^{-1}$, including Ti $t_{2g}$ and Mn $t_{2g}$ orbitals. We use $U=6$\,eV and $J=0.75$\,eV on Ti $t_{2g}$, and $U=8$\,eV and $J=0.75$\,eV on Mn $t_{2g}$. This gives an insulating solution with a band gap of $\sim 2$\,eV seen in the spectral function, as shown in  Fig.~\ref{spect-par}. The insulating state may be driven to a metallic state by reducing the value of $U$ on Mn to below $U=6$\,eV.

In the next step, in order to account for a spin-polarised solution, we first extend the Wannier basis set to include all $3d$ orbitals of Mn to allow for the high-spin state of Mn with magnetic moment of 4\,$\mu_B$. Starting from the paramagentic solutions, we introduce a spin splitting in the real part of the self energies and let the DMFT iterative cycle converge to a possibly spin-split solution. We carry out the calculations at various values of $\beta$ between 40 to 80\,eV$^{-1}$. At $\beta=40$\,eV$^{-1}$, the calculation converges still to a paramagnetic state, but when reducing the temperature we find a transition to a ferromagnetic ground state. The spectral function at $\beta=60$\,eV$^{-1}$ is shown in Fig.~\ref{het-fm-ins}. We see a clear splitting between the up and down spin channels, and a reasonably large band gap. 

What is very interesting to note is that the Ti $t_{2g}$ orbitals are now completely emptied out; the occupancies of the Ti $t_{2g}$ orbitals are negligible, whereas the occupancies of the Mn $d$ orbitals are $\sim 4.5$ each. This indicates that the electron gas
is located at the the LMO side of the interface when correlations are included and properly taken care of. In contrast, previous studies using DFT+U and hybrid functionals put the electron gas at the STO side and explain the insulating behavior of Ti electrons by a high-spin polarisation~\cite{hb-kh}. However, those approaches are inadequate when it comes to capturing the correct nature of correlations. From our calculations, however, we can show from an actual first-principles calculation the doping of Mn due to the 2DEG, arising purely from correlation effects.
Though this has been suggested before~\cite{sumilan-ch6} and used to propose a superparamagnetic state (ferromagnetic puddles in an antiferromagnetic matrix) instead of an actual intrinsic ferromagnetic insulating state, explicit calculations showing this doping of LMO did not exist. 

Next, we look at the temperature dependence of the ferromagnetic solution. Since for this purpose a larger number of calculations at different temperatures are required, we are doing calculations for a (LMO)$_{2.5}$/(STO)$_{2.5}$ heterostructure. This structure is obtained from (LMO)$_{3.5}$/(STO)$_{2.5}$ as shown in Fig.~\ref{struct} by removing one LMO layer. Hence, this reduces the number of inequivalent Mn atoms from two to one, leading to a three-impurity DMFT problem and, hence, makes calculations cheaper. Importantly, (LMO)$_{2.5}$/(STO)$_{2.5}$ shows essentially the same qualitative electronic structure as (LMO)$_{3.5}$/(STO)$_{2.5}$.

\begin{figure}
	\includegraphics[width=0.9\columnwidth]{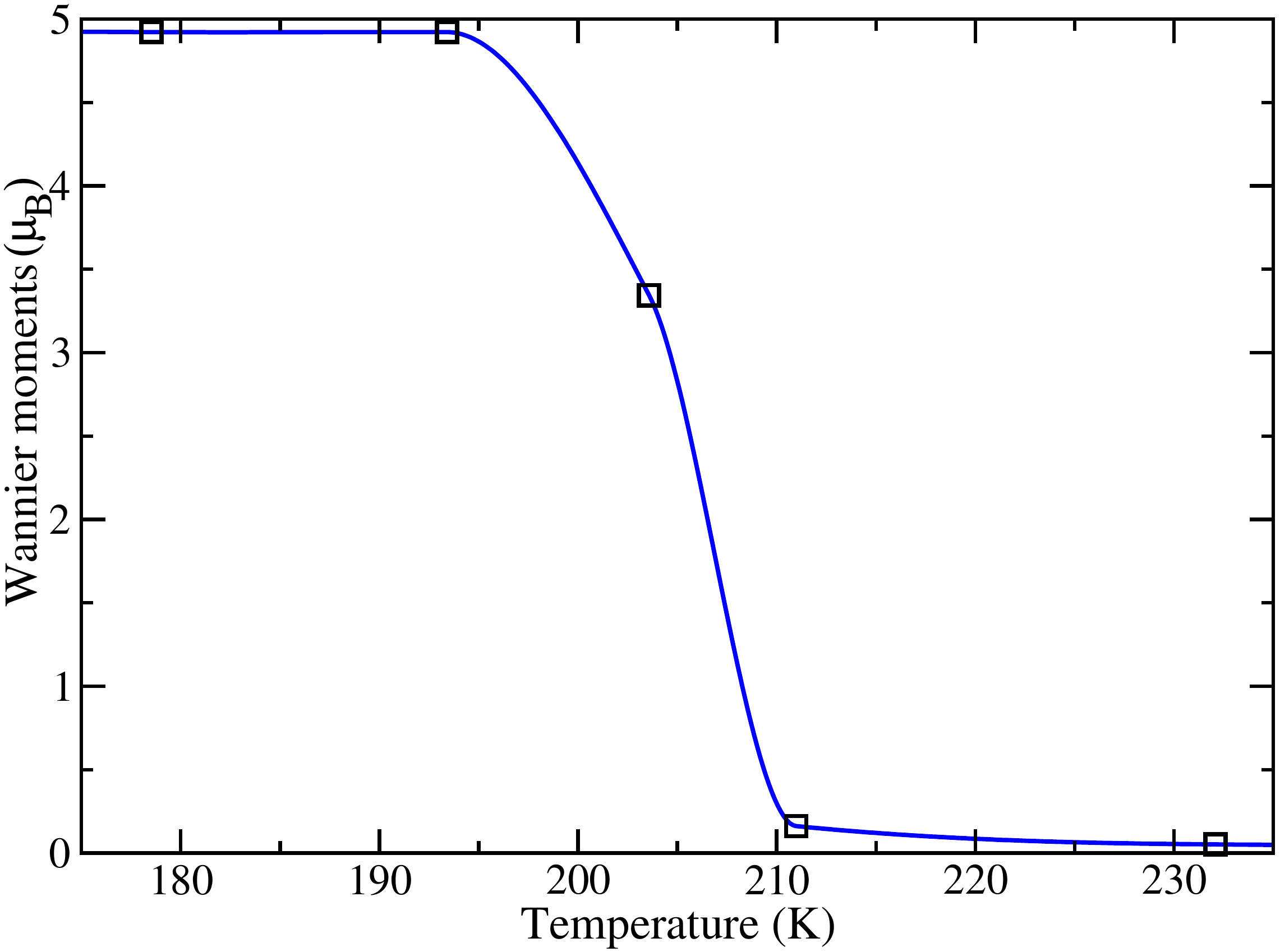}
	\caption{(Color online) Plot of Wannier magnetic moments of Mn as function of temperature.}
	\label{wann-mom}
\end{figure}
We plot the Wannier magnetic moments on Mn, obtained from the density matrix of the spin-split DMFT solution, in Fig.~\ref{wann-mom}. It is obvious that a transition from a paramagnetic to a ferromagnetic state occurs at around $\beta=60$,eV$^{-1}$, which corresponds roughly to a temperature of 200\,K. Given the fact that DMFT overestimates strongly the magnetic transition temperatures~\cite{jernej}, we would give a rough estimate for the transition temperature $T_\text{C}$ to be around 50-100\,K. A more precise prediction is beyond the capabilities of single-site DMFT calculations. We note that the N\'{e}el temperature for bulk unstrained LMO has been measured experimentally and found to be $\sim 177$\,K. We also note that our saturated magnetic moment is enhanced to $\sim 4.9$\,$\mu_B$, instead of 4\,$\mu_B$ as expected for Mn with a $d^4$ configuration. This enhancement comes from the two symmetric $n$-type interfaces giving $0.5e$ each to the 2DEG. This again confirms that indeed the electron generated due to polar catastrophe moves to the LMO side of the interface and dopes the Mn atoms instead of the Ti atoms.

\begin{figure}
	\includegraphics[width=\columnwidth]{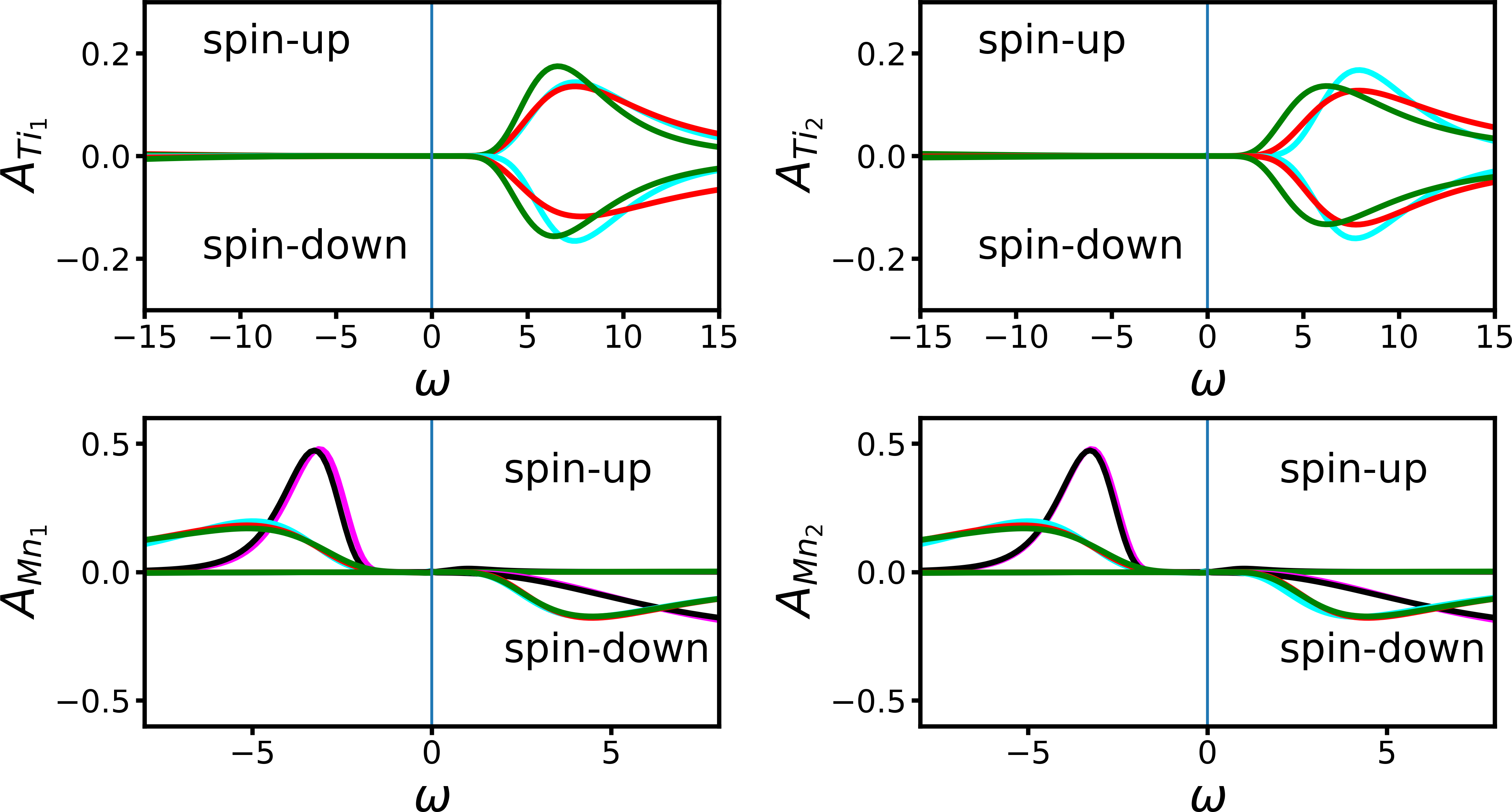}
	\caption{(Color online) DMFT spectral functions calculated starting from spin-polarised DFT, for Ti $t_{2g}$ [$d_{xy}$(cyan), $d_{yz}$(red), $d_{xz}$ (green)] and Mn $d$ [$d_{x^2-y^2}$(magenta), $d_{z^2}$ (black), and the $t_{2g}$ same as in case of Ti]. We see a FM insulating phase at $\beta=60$\,eV$^{-1}$.} 
	\label{spect-dft}
\end{figure}
\begin{figure*}[t]
	\includegraphics[width=2\columnwidth]{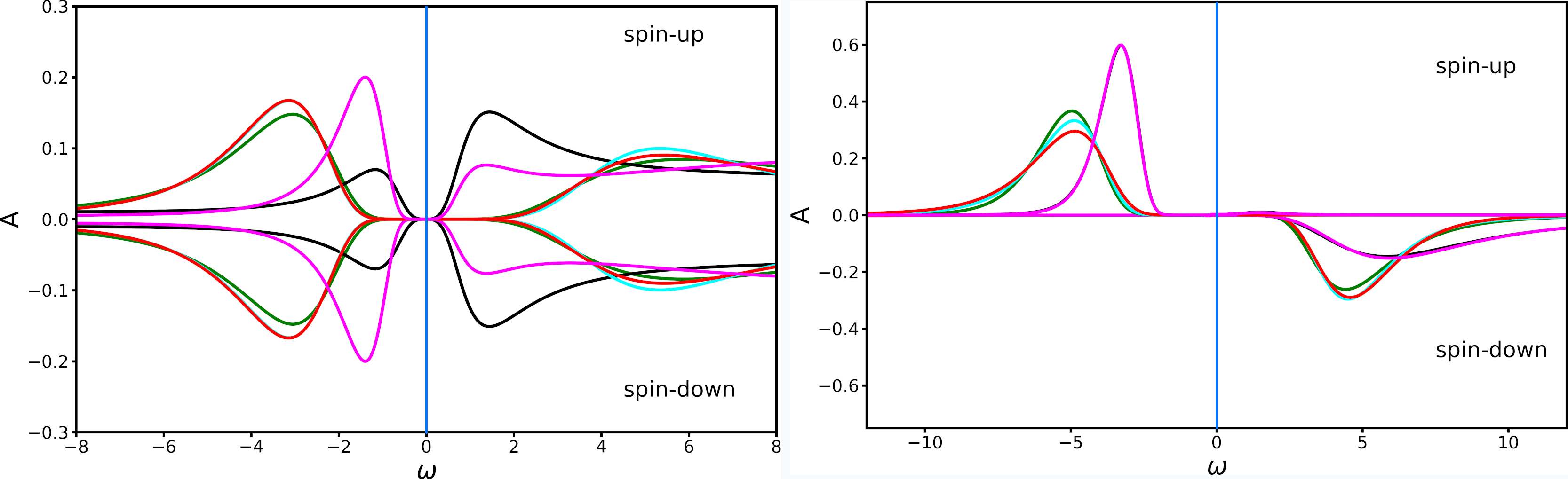}
	\caption{(Color online) Spectral functions for epitaxially strained LMO, Mn $d$ [$d_{xy}$(cyan), $d_{yz}$(red), $d_{xz}$ (green), $d_{x^2-y^2}$(magenta), $d_{z^2}$ (black)] at $\beta=40$\,eV$^{-1}$ (left panel) and $60$\,eV$^{-1}$ (right panel) respectively. At $\beta=40$\,eV$^{-1}$ we see a paramagnetic phase and at $\beta=60$\,eV$^{-1}$ we observe a ferromagnetic phase with a clear spin splitting.}
	\label{lmo-pm-fm}
\end{figure*}

Finally, to further re-affirm our results, we carry out spin-polarised DFT calculations and then look at the DMFT solution to see how correlations affect the already spin-polarised system. Our spin-polarised DFT results show Ti $d$, specifically the $d_{xy}$ states, and the Mn $t_{2g}$ states at the Fermi level. We carry out DMFT calculations at $\beta=60$\,eV$^{-1}$, within a basis set of Ti $t_{2g}$ and Mn $d$. We use the same interaction parameters as before. We find again a ferromagnetic insulating solution with a reasonable band gap as shown in Fig.~\ref{spect-dft}. Our calculations further show that also here the Ti $t_{2g}$ states empty out and the 2DEG moves to the LMO side and dopes the Mn $d$ states which become insulating. The spectral function in Fig.~\ref{spect-dft} itself looks similar to the previous case, Fig.~\ref{het-fm-ins}, where we see a clear separation between Mn $t_{2g}$ and $e_g$ states, and empty Ti $t_{2g}$ states. We thus confirm that our results are robust and independent of initial conditions.

All our calculations confirm the presence of an intrinsic ferromagnetic insulating state, arising due to electronic correlations, and clarify the location of the extra charge due to the polar catastrophe.

\paragraph{Epitaxially strained LMO}
In this final part, we want to clarify the role of strain on the magnetic state, and whether the heterostructure geometry is necessary for it. This we do by looking at bulk LMO samples.

Bulk unstrained LMO crystallizes in the orthorhombic $Pbnm$ crystal structure, and is an A-type antiferromagnetic insulator. In order to mimic the effect of epitaxial strain,
we carry out ``strained-bulk" calculations. The structural parameters ($c$ lattice parameter and
ionic positions) of the orthorhombic perovskite unit cells are 
optimized under the constraint that the two in-plane lattice vectors, defining the epitaxial 
substrate, are kept fixed to produce the specific square lattice corresponding to the substrate, in our case that of STO.
The lower panel of Fig.~\ref{struct} shows the structure of unstrained and strained LMO, viewed along the $c$ direction. We find a rather strong influence of strain on the structural parameters of LMO, particularly on the JT distortion, which becomes negligibly small in case of a compressive strain of 1.8\% generated by an STO substrate with $a_c=3.905$\,$\AA$. A strong reduction of tilt and rotation of the octahedra is also seen due to the epitaxial strain of the square substrate.

Paramagnetic DFT+DMFT calculations with interaction values $U=6$\,eV and $J=0.75$\,eV yield an insulting ground state at $\beta=40$\,eV$^{-1}$ as seen in left panel of Fig.~\ref{lmo-pm-fm}.
As for the heterstructure calculations, we then try to find a spin-split solution by introducing a spin-splitting to the paramagnetic solution. Doing so, we find a ferromagnetic insulator with a net magnetization of $M=3.96$ at $\beta=60$\,eV$^{-1}$. The spectral function for this ferromagnet is shown in the right panel of Fig.~\ref{lmo-pm-fm}. If we decrease $\beta$ slightly, there is a phase transition to a $M=0$ paramagnetic insulating phase at $\beta=50$\,eV$^{-1}$. Thus, we obtain here a very similar $T_\text{C}$ as in the previous case of the LMO/STO heterostructures.

We thus conclude that our DFT+DMFT calculations based on a five-orbital $d$ model yield a ferromagnetic insulator within a reasonable range of interaction parameters for epitaxially strained LMO as well, a phase which is not seen in the bulk unstrained condition of LMO. 
The necessary strain may be generated experimentally very easily by modern piezoelectric methods of strain generation as recently demonstrated by Hicks~\textit{et al.}~\cite{hicks}.

\paragraph{Summary and Discussion}
With the aim to provide an understanding for the ferromagnetic insulating state in LMO/STO heterostructures, we apply \textit{ab-initio} DFT+DMFT methods to both superlattices of LMO/STO and “strained-bulk” LMO structures.

We investigated the case of LMO/STO superlattice structures with comparable thicknesses of LMO and STO. The ground state was found to be insulating in both paramagnetic and ferromagnetic DFT+DMFT calculations. 
In all heterostructure geometries that we considered, the 2DEG was found to reside on the LMO side of the interface, contrary to DFT results. Though this has been suggested before, this is the first time that an actual first-principles calculation shows the doping of Mn due to the 2DEG. We also showed that the transition temperature from paramagnetic to ferromagnetic phase is high enough to be of practical relevance and accessible to experimental studies.

Finally, we showed that epitaxial straining of LMO to the square substrate of STO is key for the ferromagnetic ground state. This primarily results from a strong suppression of the JT distortion, which quenches the orbital polarization and hence antiferromagnetism in turn. For the ``strained-bulk'' structure, it was found that DFT+DMFT yields  a ferromagnetic insulating solution for small enough temperature. The critical temperature $T_\text{C}$ here is similar to the case of the  heterostructures.

Since the exotic ferromagnetic insulating state does exist in case of strained LMO, it may even be realised by using modern piezoelectric experimental techniques for generating uni/biaxial strain. Our study shows from the perspective of correlation and dynamics, the emergence of the exotic ferromagnetic insulating state, which has been touted to be immensely important in case of many spintronics applications.

\acknowledgments
This work has been funded by the Austrian Science Fund (FWF), START project Y746. Calculations have been performed on the local cluster network at TU Graz.


\begin{thebibliography}{99}
\bibitem{lao-sto} A. Ohtomo and H.Y. Hwang, Nature {\bf 427}, 423 (2004). 

\bibitem{sc-fm} S. Gariglio, N. Reyren, A. D. Caviglia, and J. -M. Triscone, J. Phys: Cond. Mat. {\bf 21}, 164213 (2009).

\bibitem{gto-sto}P. Moetakef, T. A. Cain, D. G. Ouellette, J. Y. Zhang, D. O. Klenov, A. Janotti, C. G. Van de Walle, S. Rajan, S. J. Allen, and S. Stemmer, Appl. Phys. Lett. {\bf 99}, 232116 (2011).

\bibitem{stemmer} T. A. Cain, P. Moetakef, C. A. Jackson, and S. Stemmer,  Appl. Phys. Lett. {\bf 101}, 111604 (2012).

\bibitem{banerjee-ch6} H. Banerjee, S. Banerjee, M Randeria, and T. Saha Dasgupta, Scientific Reports {\bf 5}, 18647 (2015).


\bibitem{barriocanal-ch6} J. Garcia-Barriocanal, J.C. Cezar, F.Y. Bruno, P. Thakur, N.B. Brookes, C. Utfeld, A. Rivera-Calzada, S.R. Giblin, J.W. Taylor, J.A. Duffy, S.B. Dugdale, T. Nakamura, K. Kodama, C. Leon, S. Okamoto and J. Santamaria, Nature Communications {\bf 1}, 82 (2010).

\bibitem{garcia-ch6} J. Garcia-Barriocanal, F. Y. Bruno, A. Rivera-Calzada, Z. Sefrioui, N. M. Nemes, M. Garcia-Hernández, J. Rubio-Zuazo, G. R. Castro, M. Varela, S. J. Pennycook, C. Leon, and J. Santamaria, Advanced Materials {\bf 22}, 627 (2010).

\bibitem{choi-ch6} W. S. Choi, D. W. Jeong, S. S. A. Seo, Y. S. Lee, T. H. Kim, S. Y. Jang, H. N. Lee, and K. Myung-Whun, Phys. Rev. B {\bf 83}, 195113 (2011).

\bibitem{liu2-ch6} H. M. Liu, C. Y. Ma, P. X. Zhou, S. Dong, and J.-M. Liu, J. Appl. Phys. {\bf 113}, 17D902 (2013).

\bibitem{sumilan-ch6} Y. Anahory, L. Embon, C. J. Li, S. Banerjee, A. Meltzer, H. R. Naren, A. Yakovenko, J. Cuppens, Y. Myasoedov, M. L. Rappaport, M. E. Huber, K. Michaeli, T. Venkatesan, Ariando, and E. Zeldov, Nature Communications {\bf 7}, 12566 (2016).

\bibitem{wang2015} X. R. Wang, C. J. Li, W. MThe. L\"{u}, T. R. Paudel, D. P. Leusink, M. Hoek, N. Poccia, A. Vailionis, and T. Venkatesan, Science {\bf 349}, 716 (2015).
  
\bibitem{oor} X. Zhai, L. Cheng, Y. Liu, C. M. Schlep\"{u}tz, S. Dong, H. Li, X. Zhang, S. Chu, L. Zheng, J. Zhang, A. Zhao, H. Hong, A. Bhattacharya, J. N. Eckstein, and C. Zeng, Nature Communications {\bf 5}, 4283 (2014).

\bibitem{mauger} A. Mauger A, C. Godart,  Phys Rep {\bf141}, 51, 1986.
\bibitem{katmis} J. H. Lee, et al  Nature {\bf 466}, 954, 2010

\bibitem{hou-ch6} Y. S. Hou, H. J. Xiang, and X. G. Gong, Phys. Rev. B {\bf 89}, 064415 (2014).

\bibitem{spaldin-ch6} J. H. Lee, K. T. Delaney, E. Bousquet,
N. A. Spaldin, and K. M. Rabe, Phys. Rev. B {\bf 88}, 174426 (2013).



\bibitem{hb-kh} H. Banerjee, O. Jansen, K. Held, and T. Saha-Dasgupta, Phys. Rev. B {\bf 100}, 115143 (2019)

\bibitem{blochl-ch6} P.E. Bl\"{o}chl, Phys. Rev. B  {\bf 50}, 17953 (1994).

\bibitem{kresse-ch6} G. Kresse and J. Hafner, Phys. Rev. B  {\bf 47}, 558(R) (1993).
\bibitem{kresse01-ch6}G. Kresse and J. Furthm\"{u}ller,  Phys. Rev. B  {\bf 54}, 11169 (1996).

\bibitem{pbe-ch6} J. P. Perdew, K. Burke, and M. Ernzerhof,  Phys. Rev. Letters {\bf 77}, 3865 (1996).

\bibitem{wien2k18}P. Blaha, K. Schwarz, G. K. H. Madsen, D. Kvasnicka,
J. Luitz, R. Laskowski, F. Tran, and L. D. Marks,
WIEN2k, an augmented plane wave + local orbitals
program for calculating crystal properties, (Karlheinz
Schwarz, Techn. Universität Wien, Austria) (2018).

\bibitem{triqs-parcollet} O. Parcollet, M. Ferrero, T. Ayral, H. Hafermann, I. Krivenko, L. Messio, and P. Seth, Comp. Phys. Comm. {\bf 196}, 398, (2015)

\bibitem{aichhorn1} 	M. Aichhorn, L. Pourovskii, V. Vildosola, M. Ferrero, O. Parcollet, T. Miyake, A. Georges, and S. Biermann, Phys. Rev. B {\bf 80}, 085101 (2009)

\bibitem{aichhorn2} M. Aichhorn, L. Pourovskii, and A. Georges, Phys. Rev. B {\bf 84}, 054529 (2011)

\bibitem{aichhorn3} M. Aichhorn, L. Pourovskii, P.Seth, V. Vildosola, M. Zingl, O. E. Peil, X. Deng, J. Marvlje, G. Kraberger, C. Martins, M. Ferrero, and O. Parcollet, Commt. Phys. Commun. {\bf 204}, 200 (2016)

\bibitem{werner06} P. Werner and A. J. Millis, Phys. Rev. B {\bf 74}, 155107
(2006).

\bibitem{pseth-cpc} P. Seth, I. Krivenko, M. Ferrero, and O. Parcollet, Comp. Phys. Comm. {\bf 200}, 274 (2016)

\bibitem{anisimov93} V. I. Anisimov, I. V. Solovyev, M. A. Korotin, M. T.
Czyzyk, and G. A. Sawatzky, Phys. Rev. B {\bf 48}, 16929
(1993).

\bibitem{kanamori63} J. Kanamori, Progr. Theor. Phys. {\bf 30}, 275 (1963).

\bibitem{maxent} G. J. Kraberger, R. Triebl, M. Zingl and M. Aichhorn, Phys. Rev. B {\bf 96}, 155128 (2017)


\bibitem{mellan} Thomas A. Mellan, Furio Corà, Ricardo Grau-Crespo, and Sohrab Ismail-Beigi, Phys. Rev. B {\bf 92}, 085151 (2015)






\bibitem{arima} T. Arima, Y. Tokura, and J. B. Torrance,Phys. Rev. B {\bf 48}, 17006(1993)

\bibitem{saitoh} T.  Saitoh,  A.  E.  Bocquet,  T.  Mizokawa,  H.  Namatame,  A.Fujimori, M. Abbate, Y. Takeda, and M. Takano,Phys. Rev. B {\bf 51},13942 (1995)

\bibitem{jung1} J. H. Jung, K. H. Kim, D. J. Eom, T. W. Noh, E. J. Choi, J. Yu,Y. S. Kwon, and Y. Chung,Phys. Rev. B {\bf 55}, 15489 (1997)

\bibitem{jung2} J.  H.  Jung,  K.  H.  Kim,  T.  W.  Noh,  E.  J.  Choi,  and  J.  Yu,Phys. Rev. B {\bf 57}, R11043(1998).

\bibitem{kruger} 
R.  Kr\"{u}ger,  B.  Schulz,  S.  Naler,  R.  Rauer,  D.  Budelmann,  J.Bäckström, K. H. Kim, S.-W. Cheong, V. Perebeinos, and M. R\"{u}bhausen, Phys. Rev. Lett. {\bf 92}, 097203 (2004)

\bibitem{jernej} Jernej Mravlje, Markus Aichhorn, and Antoine Georges, Phys. Rev. Lett. {\bf 108}, 197202 (2012)

\bibitem{hicks} Clifford W. Hicks, Daniel O. Brodsky, Edward A. Yelland, Alexandra S. Gibbs Jan A. N. Bruin et al,  Science  {\bf 344}, 283 (2014)































































 




\end{thebibliography}
\end{document}